\documentclass[aps,showpacs, onecolumn,twoside,floatfix,showkeys]{revtex4}
\usepackage{epsfig,amsmath,amssymb}

\newcommand{\ba}{\begin{eqnarray}}
\newcommand{\ea}{\end{eqnarray}}
\newcommand{\bmath}{\begin{mathletters}}
\newcommand{\emath}{\end{mathletters}}
\newcommand{\ban}{\begin{eqnarray*}}
\newcommand{\ean}{\end{eqnarray*}}

\begin{document}

\title{Regular and chaotic classical dynamics in the \\
U(5)-SU(3) quantum phase transition of the IBM}

\author{M. Macek}
\author{A. Leviatan}

\affiliation{Racah Institute of Physics, The Hebrew University, 
Jerusalem 91904, Israel}

\begin{abstract}
We study the classical dynamics in a generic first-order 
quantum phase transition between the U(5) and SU(3) limits 
of the interacting boson model. The dynamics is chaotic,  
of H\'enon-Heiles type, in the spherical phase and is regular, 
yet sensitive to local degeneracies, in the deformed phase.
Both types of dynamics persist in the coexistence region 
resulting in a divided phase space. 
\end{abstract}


\maketitle

The interacting boson model (IBM)~\cite{ref:Iac87} describes quadrupole 
collective states in even-even nuclei in terms of a system of $N$ 
monopole ($s$) and quadrupole ($d$) bosons, representing valence 
nucleon pairs. In addition to its traditional role of interpreting 
spectroscopic data, the model provides a fertile ground for studying 
quantum phase transitions (QPTs) and mixed regular/chaotic dynamics 
in a mesoscopic (finite) system. 
QPTs  refer to structural changes induced by a change 
of parameters $\lambda$ in the quantum Hamiltonian, $\hat{H}(\lambda)$. 
The underlying mean-field (Landau) potential, $V(\lambda)$, determines 
the nature of the QPT. In particular, 
for discontinuous (first-order) QPTs, 
$V(\lambda)$ develops multiple minima that coexist in a range of 
$\lambda$ values and cross at the critical point, $\lambda\!=\!\lambda_c$. 
In the IBM, the integrable dynamical symmetry (DS) limits relate to 
stable structural phases and QPTs are obtained by mixing terms 
from different DS chains~\cite{ref:Diep80}. 
The competing interactions that drive these QPTs can affect 
dramatically the nature of the dynamics and, in some cases, lead to the 
emergence of quantum chaos.

QPTs~\cite{ref:Diep80,ref:Cejnar09,ref:Iac11} 
and chaos~\cite{ref:Whel93,ref:Mace07}
have been studied extensively in the IBM, albeit, with a simplified 
Hamiltonian which, for first-order QPTs, gave rise   
to extremely low barrier and narrow coexistence region. 
Recently, the identification of IBM Hamiltonians without such 
restrictions~\cite{ref:Lev06}, enabled
a comprehensive analysis of generic first-order 
QPTs between spherical and deformed 
shapes~\cite{ref:MacLev11,ref:LevMac12,ref:MacLev12}.
The dynamics inside the phase coexistence region was found to exhibit 
a very simple pattern. 
A~classical analysis revealed a robustly regular dynamics confined to 
the deformed region and well separated from a chaotic dynamics 
ascribed to the spherical region. This divided phase space structure 
manifests itself also in the quantum analysis, disclosing regular 
rotational bands in the deformed region amidst a complicated environment.
In the present contribution, we illuminate the origin of this intricate 
interplay of order and chaos for a particular case, where the stable 
spherical and deformed phases posses U(5) and SU(3) DS, respectively.  

Apart from kinetic rotational terms, the relevant Hamiltonian 
(up to a scale) is 
\ba
\label{eq:Hint}
\hat{H}_1(\rho) &=& 
2(1 - 2\rho^2)\hat{n}_d(\hat{n}_d - 1) + 
2 R^{\dag}_2(\rho) \cdot \tilde{R}_2(\rho)  ~,
\label{eq:H1} \\
\hat{H}_2(\xi) &=& 
\xi P^{\dag}_0 P_0 + P^{\dag}_2 \cdot \tilde{P}_2 ~.
\label{eq:H2} 
\ea
Here $\hat{n}_d$ is the $d$-boson number operator, 
$R^{\dag}_{2\mu}(\rho) \!=\! \sqrt{2}s^\dag d^\dag_\mu + 
\rho\sqrt{7}(d^\dag d^\dag)^{(2)}_\mu$, 
$P^{\dag}_0 = d^\dag \cdot d^\dag - 2 (s^\dag)^2$, 
$P^{\dag}_{2\mu} \!=\! 2 s^\dag d^\dag_\mu + 
\sqrt{7}(d^\dag d^\dag)^{(2)}_\mu$ 
and the centered dot implies a scalar product.
The parameters that control the QPT 
are $\rho$ and $\xi$, 
with $0\leq\rho\leq 1/\sqrt{2}$ and $\xi\geq 0$.
For $\rho=0$ and $\xi=1$, one recovers the 
U(5) and SU(3) DS limits, where the Hamiltonians 
\ba
\hat{H}_{1}(\rho=0) &=& 2\hat{n}_{d}[2\hat{N}-\hat{n}_d -1] ~,
\label{eq:Hu5}
\\ 
\hat{H}_{2}(\xi=1) &=& [-\hat{C}_{\rm SU(3)} + 2\hat{N}(2\hat{N}+3)] ~,
\label{eq:Hsu3}
\ea
involve the relevant Casimir operators. 
The two Hamiltonians of Eqs.~(\ref{eq:H1})-(\ref{eq:H2}) 
coincide at the critical point 
$\rho_c\!=\! 1/\sqrt{2}$ and $\xi_c \!=\!0$: 
$\hat{H}_{1}(\rho_c) \!=\! \hat{H}_{2}(\xi_c)$. 

The classical limit is obtained through the use of coherent 
states, rescaling and taking $N\rightarrow\infty$, with 
$1/N$ playing the role of $\hbar$. The derived classical Hamiltonians, 
$\mathcal{H}_1(\rho)$, $\mathcal{H}_2(\xi)$, 
involve complicated expressions of shape variables $(\beta,\gamma)$, 
Euler angles and their conjugate momenta. 
Setting the latter to zero, yields the following classical potentials
\ba
\label{eq:V}
V_1(\rho) &=& 
2\beta^2  - \beta^4 /2 
- 2 \rho \sqrt{2 - \beta^2} \beta^3\cos3\gamma ~,
\label{eq:V1}\\
V_2(\xi) &=& 
\xi\left(4 - 6\beta^2 + 9\beta^4/4\right) 
+ 2\beta^2 - \beta^4/2
- \sqrt{4 - 2\beta^2} \beta^3\cos3\gamma ~.
\label{eq:V2}
\ea
$V_1(\rho)$ and $V_2(\xi)$ serve as the Landau potentials with 
the equilibrium deformations 
$(\beta_{\mathrm{eq}},\gamma_{\mathrm{eq}})$ as order parameters. 
The potential $V_1(\rho)$ [$V_2(\xi)$] has a global spherical [deformed] 
minimum with, respectively, $\beta_{\mathrm{eq}}\!=\!0$ 
[$\beta_{\mathrm{eq}}\!=\!2/\sqrt{3},\gamma_{\mathrm{eq}}\!=\!0$]. 
At the spinodal point ($\rho^{*}\!=\!1/2$), $V_1(\rho)$ develops an additional 
local deformed minimum. The two minima become degenerate at the 
critical point $\rho_c\!=\!1/\sqrt{2}$ (or~$\xi_c\!=\!0$), 
and are separated by a barrier of height $V_b \!=\!0.268$ 
(compared to $V_b \!=\!0.0018$ in previous 
works~\cite{ref:Whel93}). The spherical minimum 
turns local in $V_2(\xi)$ for $\xi>\xi_c$ and disappears at the anti-spinodal 
point~($\xi^{**}\!=\!1/3$). 
The  order parameter ${\beta_{\mathrm eq}}$, 
is a double-valued function 
in the coexistence region (in-between $\rho^{*}$ and $\xi^{**}$) 
and a step-function outside it.

\begin{figure}
\includegraphics[height=0.485\textwidth]{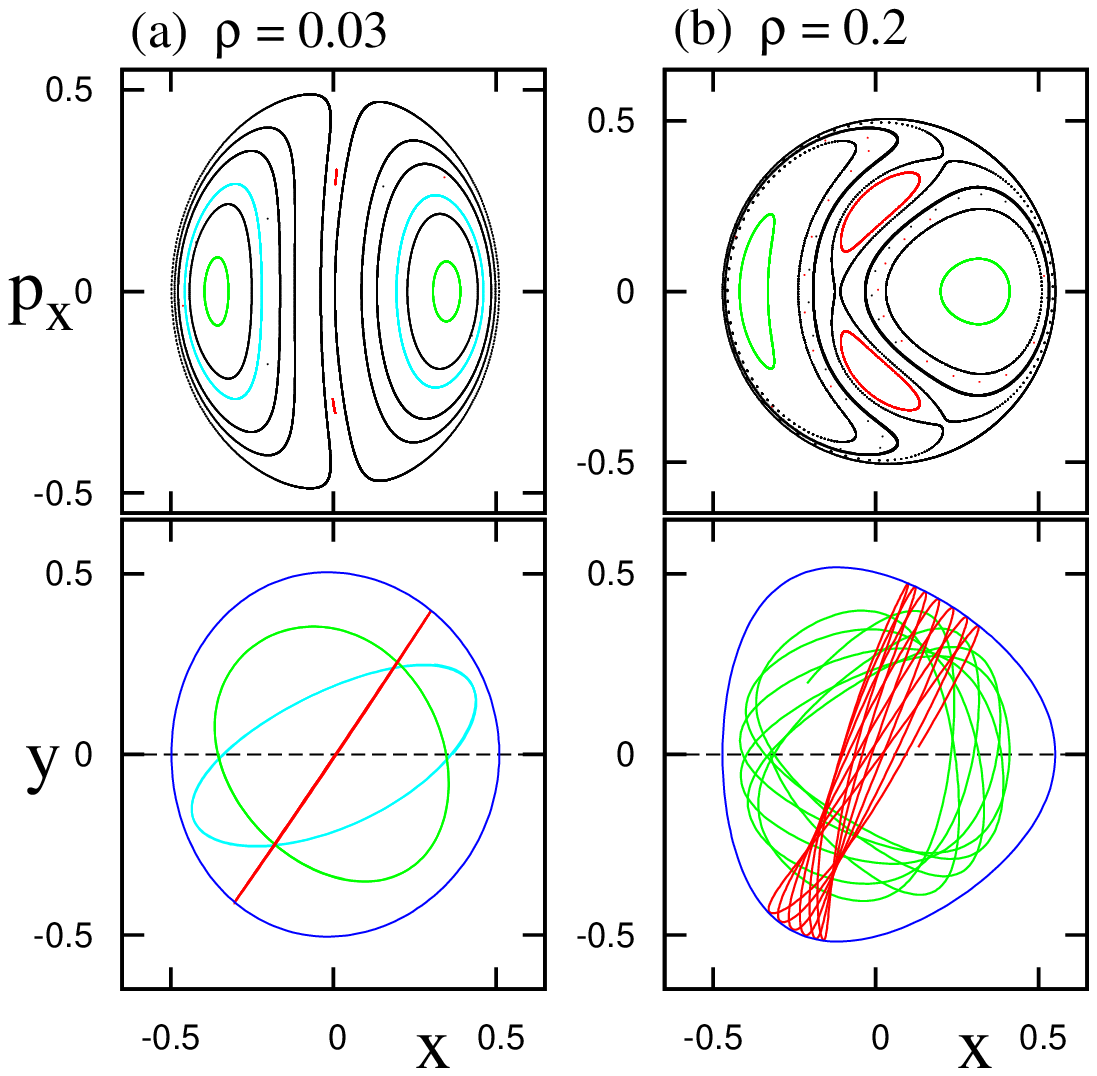}
\includegraphics[height=0.485\textwidth]{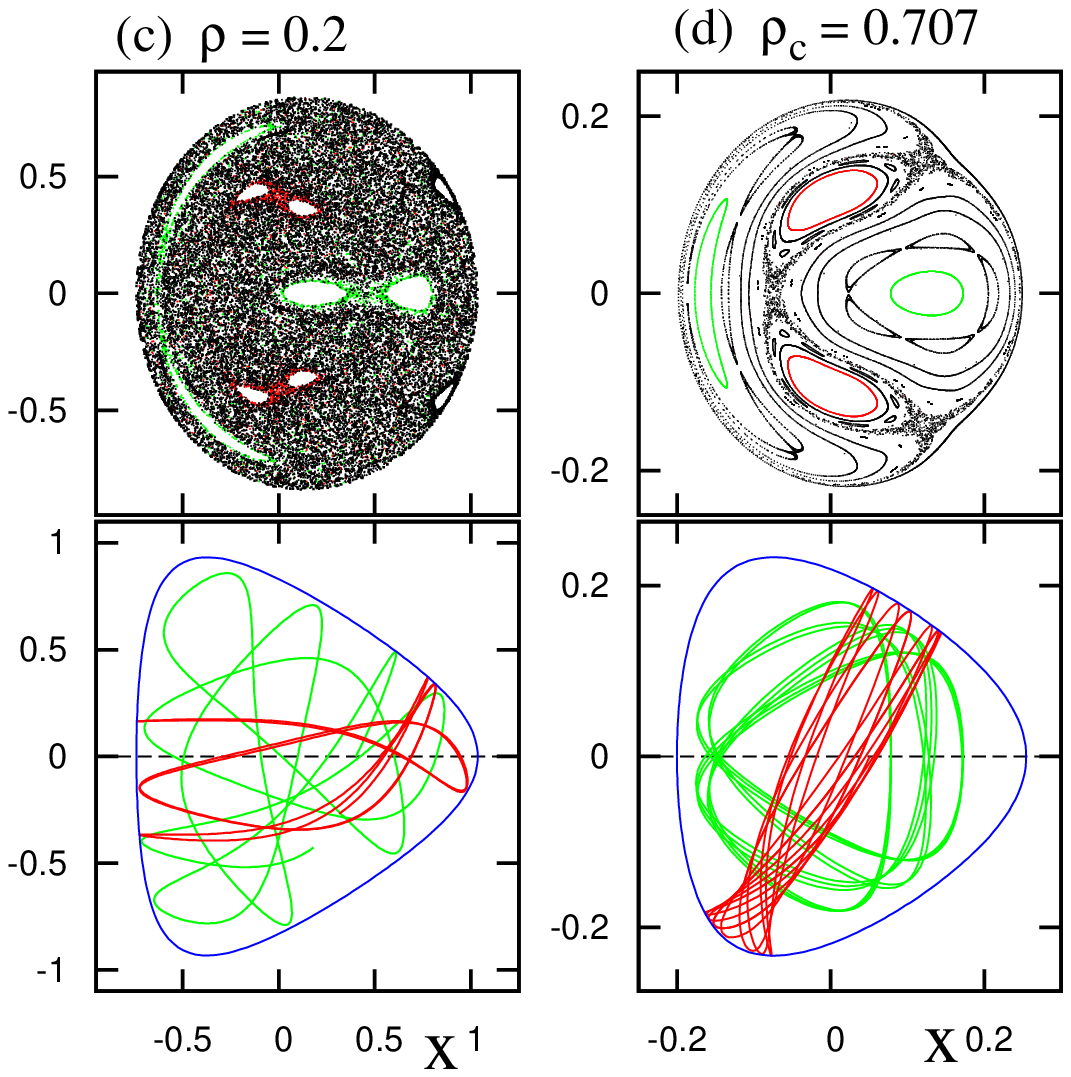}
\caption{Poincar\'e sections (top row), plotted at $y=0$, 
and selected trajectories (bottom row), 
depicting the classical dynamics of $\hat{H}_{1}(\rho)$~(\ref{eq:H1}) 
for several values of $\rho$ and energies $E$.
(a)~$\rho=0.03$, $E=E_{1}/21$. 
(b)~$\rho=0.2$, $E=5E_{1}/12$.
(c)~$\rho=0.2$, $E= 12E_{1}/21$.
(d)~$\rho_c=1/\sqrt{2}$, $E=E_{1}/12$ (the critical point).
The energies are given with respect to the domain boundary
$E_1\equiv V_{1}(\rho;\beta=\sqrt{2})=2$.}
\label{fig:I}   
\end{figure}

\begin{figure}
\includegraphics[height=0.475\textwidth]{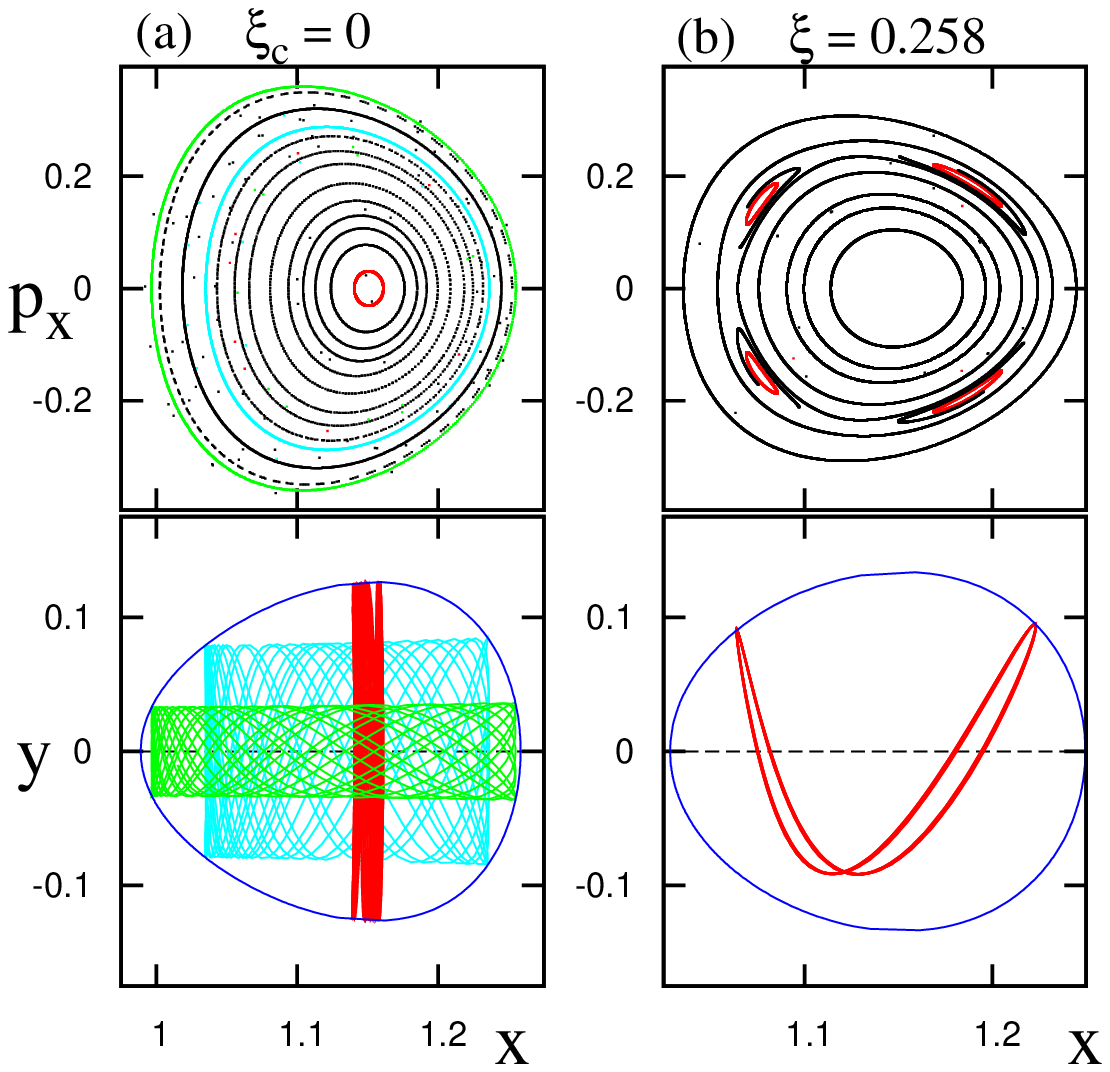}
\includegraphics[height=0.475\textwidth]{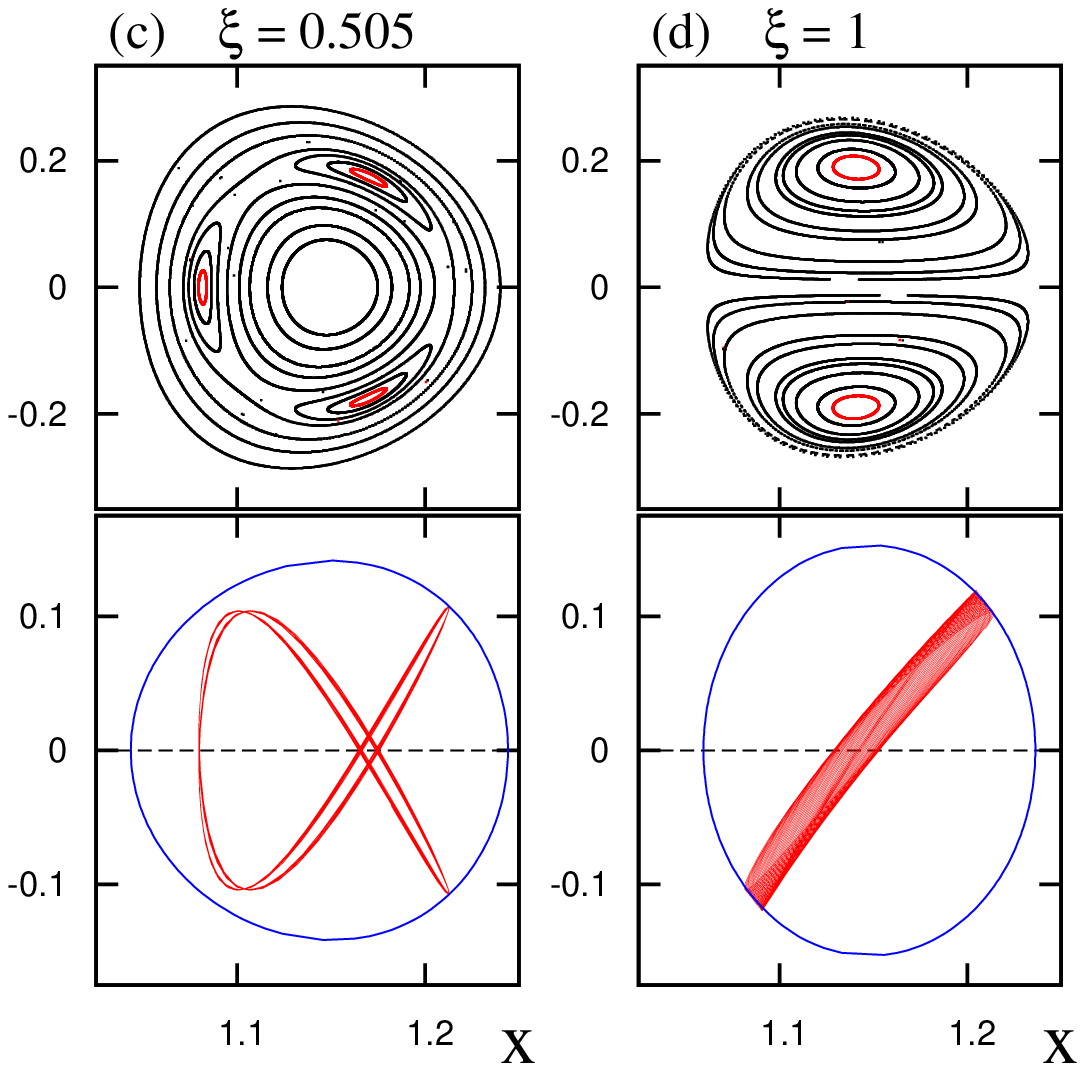}
\caption{Same as in Fig.~\ref{fig:I} but for the classical 
dynamics of $\hat{H}_{2}(\xi)$~(\ref{eq:H2}) for selected values of 
$\xi$ and energy $E=E_2/21$, where 
$E_2\!=\! V_{2}(\xi;\beta\!=\!\sqrt{2}) \!=\! 2\!+\!\xi$.
Panel~(a) shows a typical pattern encountered for most values of 
$\xi\geq0$. Panels (b)-(c)-(d) portray the surfaces in the presence of 
local degeneracies of normal modes, $\epsilon_{\beta}/\epsilon_{\gamma} = 
1/2,\,2/3,\,1$, respectively. Panel~(d) corresponds to the SU(3) DS 
limit, Eq.~(\ref{eq:Hsu3}).}
\label{fig:II}   
\end{figure}

The classical analysis simplifies considerably when the dynamics 
is restricted to $L=0$ vibrations. In this case, the classical 
Hamiltonians, $\mathcal{H}_1(\rho)$, $\mathcal{H}_2(\xi)$, 
become two-dimensional in the polar coordinates  
$\beta\in[0,\sqrt{2}]$, $\gamma\in[0,2\pi)$ 
and momenta $p_\beta\in[0,\sqrt{2}]$, $p_\gamma\in[0,1]$ 
(or, equivalently, in Cartesian coordinates, 
$x\!=\!\beta\cos{\gamma}$, $y\!=\!\beta\sin{\gamma}$ and $p_x$, $p_y$). 
The classical motion can then be depicted conveniently via 
Poincar\'e surfaces of section~\cite{ref:Gutz90}, 
shown at prescribed energies 
in Figs.~\ref{fig:I}-\ref{fig:II}, along with selected trajectories.

The classical dynamics on the spherical side of the QPT 
($0\leq\rho\leq\rho_c$), is governed by $\mathcal{H}_1(\rho)$. 
In the U(5) limit ($\rho=0$), the system is integrable and 
$\mathcal{H}_1(\rho=0) = 2(T + \beta^2) - (T + \beta^2)^2/2$, 
where $T=p_\beta^2 + \beta^{-2} p_\gamma^2$. 
As shown in Fig.~\ref{fig:I}(a), 
the sections, for small $\rho$, 
show the phase space 
portrait typical of a weakly perturbed anharmonic (quartic) oscillator 
with two major regular islands and quasi-periodic trajectories. 
The effect of increasing $\rho$ on the dynamics in the vicinity 
of the spherical minimum ($x\approx 0$), can be inferred from a small 
$\beta$-expansion of the potential, 
$V_1(\rho)\approx 2\beta^2 - 2 \rho \sqrt{2} \beta^3\cos3\gamma$. 
To this order, $V_1(\rho)$ coincides with the well-known 
H\'enon-Heiles (HH) potential~\cite{ref:Heno64}.
As shown for $\rho\!=\!0.2$, at low energy [Fig.~\ref{fig:I}(b)],
the dynamics remains regular, and two additional islands show up. 
At higher energy [Fig.~\ref{fig:I}(c)], one observes a marked onset 
of chaos and an ergodic domain. 
The energy where the order-to-chaos transition occurs decreases with $\rho$.
This typical HH-type of behavior persists
in the vicinity of the spherical minimum 
throughout the coexistence region, including the critical point 
[Fig.~\ref{fig:I}(d)]. It is also present in the region where 
the spherical minimum is only local ($0\leq\xi\leq\xi^{**}$), since
$V_2(\xi)\approx 4\xi + (2 - 6\xi)\beta^2 - 2 \beta^3\cos3\gamma$. 

The classical dynamics on the deformed side of the QPT 
($\xi_c\leq\xi\leq 1$), is governed by $\mathcal{H}_2(\xi)$ 
and has a very different character, being robustly regular 
in the vicinity of the deformed minimum ($x\approx 1$). 
At low energy, the motion reflects the $\beta$ and $\gamma$ normal 
mode oscillations about the minimum. 
As shown in  Fig.~\ref{fig:II}(a), the trajectories form a single set 
of concentric loops around a single stable (elliptic) fixed point. 
They portray $\gamma$-vibrations at the center 
of the surface ($p_x\approx 0$) and $\beta$-vibrations at the perimeter 
(large $\vert p_x\vert$). 
This regular pattern of the dynamics is found for most values of 
$\xi\geq 0$ both inside and outside the phase coexistence region.
Noticeable exceptions occur in the presence of 
resonances, which appear when the ratio of normal mode frequencies, 
$R\equiv \epsilon_\beta/\epsilon_\gamma$, is a rational number. 
At low energy, this happens at discrete values of the control parameter 
$\xi\approx\xi_R$, in a narrow interval around $\xi_R = (3R-1)/2$.
Panels (b)-(c)-(d) of Fig.~\ref{fig:II}
show examples of such scenario for $R=1/2,\,2/3,\,1$. The corresponding 
surfaces exhibit four, three and two KAM islands, respectively. 
The phase space portrait for ($\xi=1,R=1$), shown in Fig.~\ref{fig:II}(d), 
corresponds to the integrable SU(3) DS limit, Eq.~(\ref{eq:Hsu3}), 
and is the same for any energy. Interestingly, in the coexistence region, 
where the Landau potential accommodates both the spherical and deformed 
minima, each minimum preserves its own characteristic dynamics.
This is evident in Fig.~\ref{fig:I}(d) and Fig.~\ref{fig:II}(a), which 
depict the dynamics of the same system 
[$\hat{H}_{1}(\rho_c)=\hat{H}_{2}(\xi_c)$] at the same energy, 
but in different regions ($x$-ranges) of phase space.

In summary, we have shown the distinct morphology of classical orbits and 
different onset of regularity and chaos in a generic (high-barrier) 
first-order QPT between the U(5) [spherical] 
and SU(3) [deformed] phases of the IBM. The spherical phase displays a 
chaos-susceptible dynamics, similar to the H\'enon-Heiles system, 
due to a $\beta^3\cos 3\gamma$ perturbation, 
significant near the spherical minimum. In contrast, the low energy 
dynamics in the deformed phase is robustly ordered, and 
reflects the $\beta$-$\gamma$ vibrations about 
the deformed minimum. This regular dynamics is sensitive 
to local degeneracies of these normal modes. 
The two types of dynamics preserve their identity and can be detected 
as long as the relevant minimum exists. 
This leads to a divided phase space structure 
throughout the phase-coexistence region, 
as found in~\cite{ref:MacLev11,ref:LevMac12,ref:MacLev12}. 

This work is supported by the Israel Science Foundation. M.M. acknowledges 
support by the Golda Meir Fellowship Fund and the Czech Ministry of 
Education (MSM 0021620859).

\end{document}